\documentclass[10pt,a4paper]{article}

\usepackage{graphicx}
\usepackage{hyperref} 
\usepackage{amsmath,amsfonts}%
\usepackage{tabularx}
\usepackage{enumerate}
\usepackage{algorithm}
\usepackage{algpseudocode}
\usepackage{xspace}
\usepackage{authblk}
\usepackage{multirow}

\usepackage{longtable}
\usepackage{graphicx}
\usepackage[colorinlistoftodos]{todonotes}
\usepackage{subfigure} 
\usepackage[utf8]{inputenc}

\usepackage{graphicx}
\usepackage[colorinlistoftodos]{todonotes}
\usepackage{tikz}   

\graphicspath{{Figures/}} 

\def\D{\mathcal{D}}

\def\Dg{\mathcal{D}_{\gamma}}

\newenvironment{psmallmatrix}
  {\left[\begin{smallmatrix}}
  {\end{smallmatrix}\right]}

\providecommand{\keywords}[1]
{
  \small	
  \textbf{\textit{Keywords---}} #1
}

\definecolor{lime}{HTML}{A6CE39}
\DeclareRobustCommand{\orcidicon}{%
	\begin{tikzpicture}
	\draw[lime, fill=lime] (0,0) 
	circle [radius=0.16] 
	node[white] {{\fontfamily{qag}\selectfont \tiny ID}};
	\draw[white, fill=white] (-0.0625,0.095) 
	circle [radius=0.007];
	\end{tikzpicture}
	\hspace{-2mm}
}

\foreach \x in {A, ..., Z}{%
	\expandafter\xdef\csname orcid\x\endcsname{\noexpand\href{https://orcid.org/\csname orcidauthor\x\endcsname}{\noexpand\orcidicon}}
}

\title{Rao-Burbea centroids applied to the statistical characterisation of time series and images through ordinal patterns}

\author[1,3,4,*] {Diego M. Mateos\orcidA{}}
\author[2]{Leonardo E. Riveaud }
\author[1,5]{Pedro W. Lamberti}

\affil[1]{\normalsize Consejo Nacional de Investigaciones Cient\'ificas y T\'ecnicas (CONICET), Argentina.}
\affil[2]{\normalsize Facultad de Ingenier\'ia, Universidad Nacional del Comahue (FAIN, UNComa)}
\affil[3]{\normalsize Facultad de Ciencia y Tecnolog\'{\i}a. Universidad Aut\'{o}noma de Entre R\'{\i}os (UADER). Oro Verde, Entre R\'{\i}os, Argentina.}
\affil[4]{\normalsize Instituto de Matem\'{a}tica Aplicada del Litoral (IMAL-CONICET-UNL), CCT CONICET, Santa F\'e, Argentina.}
\affil[5]{\normalsize Facultad de Matem\'atica Astronom\'ia, F\'isica y Computaci\'on (FaMAF), Universidad Nacional de C\'ordoba. C\'ordoba, Argentina.}

\affil[*]{Corresponding author: Diego M. Mateos, mateosdiego@gmail.com.}
\begin{document}

\date{}
\maketitle

\begin{abstract}

Divergences or similarity measures between probability distributions have become a very useful tool for studying different aspects of statistical objects such as time series, networks and images. Notably not every divergence provides identical results when applied to the same problem. Therefore it is convenient to have the widest possible set of divergences to be applied to the problems under study. Besides this choice an essential step in the analysis of every statistical object is the mapping of each one of their representing values into an alphabet of symbols conveniently chosen. In this work we attack both problems, that is, the choice of a family of divergences and the way to do the map into a symbolic sequence. For advancing in the first task we work with the family of divergences known as the Burbea-Rao centroids (BRC) and for the second one we proceed by mapping the original object into a symbolic sequence through the use of ordinal patterns. Finally we apply our proposals to analyse simulated and real time series and to real textured images. The main conclusion of our work is that the best BRC, at least in the studied cases, is the Jensen Shannon divergence, besides the fact that it verifies some interesting formal properties.

\end{abstract}


\vspace{1cm}
\textbf{The notion of distinguishability is of crucial importance in probability theory and in general in every statistical theory. Due to statistical fluctuations it becomes difficult to distinguish between close probability distributions. This problem involves two issues. The first one is the definition of distance measures between probability distributions and the second one is the introduction of distinguishability criteria. On the other hand, many physical phenomena can be represented by means of a time series, which could be generated by chaotic dynamics or by a random proccess. It is necessary, in order to be able to study these dynamics, to make a correspondence between the values of the time series under study with a symbolic sequence. In several occasions  this task is the most difficult to perform. In this work we address the different problems indicated above. First, we study the applicability of a family of divergences between probability distributions, known as Burbea-Rao centroids, to the study of time series. We then investigate the use of ordinal patterns as a way to assign a symbolic sequence to a time series. With these two tools we study the statistical characteristics of simulated and real world time series as well as to textured 2D images. As the Burbea-Rao centroids depend on a concave functional we test the behaviour of this divergences when we change it. Among them one of the choosen funtional leads to the Jensen Shannon divergence. In addition of demonstrating the robustness of the scheme presented it was possible to conclude that in all the investigated examples the best results correspond to the Jensen Shannon divergence.}

\section{Introduction}

There exists a  profusion of divergences between probability distributions. Remarkably when some of them are applied to the same statistical problem, in general they do not lead to indistinguishable results. Therefore, might be useful to have a large  set of divergences. In general the divergences have different origins. Some are statistical, others originated in information theory and others borrowed from the realm of pure mathematics. The Fisher's metric is a conspicuous example of the first kind \cite{fisher1925theory}; and the Kullback-Leibler is a well known example originated in information theory \cite{cover2012elements}. Measures of similarity (or dissimilarity), between probability distributions have become of great interest in  physics (classical and quantum), biology and many other areas of science and technology \cite{parry2000probability,ondimu2008effect,kostin2010probability,majtey2005jensen}.

In the realm of statistical physics particularly notable has been the use of divergences in the context of out-of-equilibrium systems. The obligatory reference on this subject is the work of G. Crooks and D. Sivak. There these authors give a physical interpretation to different measures of divergence when are applied to the study of conjugate ensembles of non equilibrium trajectories. The relative entropy is related to the dissipation, the Jeffreys divergence is the average dissipation of the forward and reverse evolution (hysteresis), the Jensen – Shannon divergence has been proposed as a magnitude of the time arrow, and the Chernoff divergence is the work cumulant generating function \cite{crooks}.

As it was said before having a variety of divergence could be useful for both pure theoretical studies and in the context of applications. Sometimes it is possible to introduce families of divergences, labelling each member with a parameter \cite{osan2018monoparametric,majtey2004monoparametric} or by giving a general structure depending on certain functional with adequate characteristics. An example of this last case are the Csiszar's divergences or $f$-divergences. 

Let $\Omega = \{\omega_1,...,\omega_{N+1}\}$ be a discrete sample space. The set of probability distributions on $\Omega$ can be identified with the simplex $\mathbb{P}^N=\{x^i \geq 0, i=1,...,N+1, \sum_{i=1}^{N+1} x^i=1\}$. Given two probability distributions $P=\{p_i\}_{i=1}^{N+1}$ and $Q=\{q_i\}_{i=1}^{N+1}$ belonging to $\mathbb{P}^N$, a Csiszar's divergence is defined by \cite{csiszar1974information}
\[
\mathcal{D}_f(P,Q)= \sum_{i=1}^{N+1} f \left( \frac{p_i}{q_i}\right) q_i
\] 
where $f(x)$ is a convex function such that $f(1)=0$. This family has been extensively studied in the context of information geometry \cite{burbea1982entropy}. A remarkable result is that when $p_i$ and $q_i=p_i + \delta p_i$ are two close probability distributions, the divergence $\mathcal{D}_f(P,Q)$ is proportional to the Fisher (Riemannian) metric:
\begin{equation}
\label{ec:aprox}
\mathcal{D}_f(P,P+\delta P) \sim \frac{f''(1)}{2} \sum_i \frac{\delta p_i^2}{p_i}
\end{equation}

The above mentioned Kullback-Leibler divergence corresponds to $f(t)=t ~ \log t$ and it is not symmetric and the Jensen-Shannon divergence (JSD) it is obtained by taking $f(t)= (t+1) \log(\frac{2}{t+1}) + t ~ \log t$, which clearly is symmetric \cite{ali1966general}. 

For the study of dynamical phenomena, we need to have a sequence of measurements related to them. These sequences are usually given in the form of time series which allow extracting information on the underlying physical system under study. Something similar happen with images, where we have a two-dimensional matrix which contains the information about each pixel of the images. The  time series or images can be associated with a probability distribution function (PDF) and, from this to have a way for applying the divergences.  To this aim it is necessary to map the time series or images in a finite alphabet. There exist many methods to discretize a continuous time series, such as binarization or the wavelet analysis. However these methods do not take into account the relation between the value for a given time and the neighbouring values. In 2002 C. Bandt and B. Pompe introduced the ordinal patterns for time series discretization  \cite{Bandt2002}. This approach has been extensively used to different problems with very relevant results. Its main advantage is that reveals the underlying dynamics in the process that generates the time series. For a review of these properties we suggest the work of  M. Zanin and F. Olivares in which the ordinal pattern  approach is studied in detail  \cite{zanin2021ordinal}.

Here we use the Burbea-Rao centroids (BRC) as the dissimilarity measures and we build a symbolic alphabet formed by ordinal patterns, in which the original time series or the 2D images are mapped. Then we apply an segmentation procedure meaning this a way to detect dynamical changes in the corresponding symbolic sequence. All this scheme is used to the study of the statistical properties of simulated time series and to real 2D images.  

This manuscript is organised as follows. In  section \ref{sec:gamma} we review some properties of the BRC. We show that the BRC can be thought as a deformations of the Euclidean metric and we indicate a particular character of the JSD as a BRC. In section \ref{sec:op_gammaD} we give a brief explanation of the ordinal patterns and how them can be evaluated in some concrete cases. In  section \ref{sec:application} we combine the use of the BRC, really a particular case that we called the $\gamma$-divergences, with the ordinal patterns to four examples, three related to time series analysis and one to images analysis. Finally, in  section \ref{sec:Discussion} we discussed our results as well as some other possible frameworks of applicability of our scheme.

\section{Burbea-Rao centroids}
\label{sec:gamma}
In 1982 J. Burbea and C.R. Rao introduced a family of divergences now known as the Burbea-Rao centroids (BRC). They started from a concave $\Phi$ functional. A generic member of this family is given by:
\begin{equation}
    \mathcal{J}_\Phi(P,Q) = \frac{1}{2} [\Phi(P) + \Phi(Q)] - \Phi\left(\frac{P+Q}{2}\right) \label{BRC}
\end{equation}
where $P$ and $Q$ are two discrete probability distributions belongings to $\mathbb{P}^N$.

Incidentally it is worth mentioning that a centroid can be interpreted as a deformation of the square of the Euclidean metric. Indeed, let us write the square of the Euclidean distance between $P$ and $Q$ in the form:
\begin{equation}
\label{ec:euclidea}
\mathcal{E}(P,Q)=\mathop{\sum}_{i}(q_{i}-p_{i})^{2}= \mathop{\sum}_{i} \left( 2~q_{i}^{2}+2~p_{i}^{2}-(p_{i}+q_{i})^{2}\right)
\end{equation}
which can be rewritten as a BRC with $\Phi = \sum_i x_i g(x_i)$ and $g(x) =x$. In this sense we can think that a BRC is a ``deformation'' of the Euclidean metric through the map $\sum_i x_i^2 \rightarrow \Phi$. Analogously the Jensen-Shannon divergence (JSD) mentioned above can be written as a BRC with $\Phi = \sum_i x_i g(x_i)$ and $g(x) = \frac{1}{2} \log x$. The explicit expression of the JSD is:

\begin{equation}
    \mathcal{D}_{JS}(P,Q) = \sum_i \frac{1}{2} p_i \log p_i + \frac{1}{2} q_i \log q_i - \frac{1}{2}(p_i+q_i) \log\left(\frac{p_i + q_i}{2} \right)
\end{equation}

An important characteristic of the JSD is that it is the only BRC that is a Csiszar divergence. This can be easily proved by doing a Taylor expansion of (\ref{BRC}) up to second order. Not every BRC is a metric in the sense that verifies the triangle inequality. However it has been shown that the square root of the JSD is a true metric for the simplex $\mathbb{P}^N$ as well as the Euclidean metric (\ref{ec:euclidea}). In general the metric character of a BRC will depend on the functional $\Phi$.

One advantage of a BRC that is particularly interesting for us is that by modifying the function $\Phi$, we can determine the divergence that more adequately allows to characterise the statistical object under study.  

\vspace{2mm}

\subsection{Weighted BRC}

In 1991 J. Lin introduced a generalised version of the JSD by assigning different weights to each probability distribution, $P$ and $Q$. Let $\pi_P$ and $\pi_Q$ two non negative numbers such that $\pi_P + \pi_Q=1$. Then the weighted JSD is given by:
\begin{equation}
   \mathcal{D}_{JS}(P,Q; \pi_P,\pi_Q)= H_S(\pi_P P + \pi_Q Q) - \pi_P H_S(P) - \pi_Q H_S(Q) 
\end{equation}
with $H_S(P) \equiv -\sum_i p_i \log(p_i)$ the Shannon entropy. We can proceed in the same way with a generic BRC and define the weighted BRC as follows:
\begin{equation}
    \mathcal{J}_\Phi(P,Q,\pi_P,\pi_Q) \equiv \pi_P \Phi(P) + \pi_Q \Phi(Q) - \Phi(\pi_P P + \pi_Q Q) \label{wBRC}
\end{equation}
Due to the functional $\Phi$ it is assumed to be concave, as a consequence of the Jensen inequality, the quantity (\ref{wBRC}) is always non negative.

Another possible generalisation of the BRC is to evaluate it among several probability distributions each one with different weight.  To fix ideas let us suppose $K$ probability distributions $P^{(1)},...,P^{[K)}$ and weights $\pi_1,...\pi_K$. We can define a weighted RBC among these probability distributions in the form:

\begin{equation}
\label{weighted}
\mathcal{J}_{\Phi}(P^{(1)},...,P^{(K)};\pi_1,...,\pi_N) \equiv \sum_{k=1}^K \pi_k \Phi(P^{(k)}) - \Phi\left(\sum_{k=1}^K \pi_k P^{(k)}\right)  
\end{equation}
Again due to the concavity of $\Phi$ and the Jensen inequality this last quantity is non negative.

\section{Ordinal Patterns and Burbea Rao centroids}
\label{sec:op_gammaD}

As was remarked in the introduction for studying the statistical properties of a time series, and with the purpose of assigning probability distributions to it, it is indispensable to map the original data set, that is, the values of the time series, into symbolic sequences. There are different methods in the literature for doing this, such as the discretization, binarization and  wavelet's histograms, among others. An alternative method proposed by Bandt and Pompe (BP) \cite{Bandt2002} consists in transforming the time series, via a non parametric transformation, into a sequence of patterns and then making inference over these patterns. With this, the analysis gains robustness and becomes apt to detect relevant causal information related to the unobserved variables that control the underlying dynamics of the system.
The BP approach is based on the computation of the Shannon entropy from the histogram of causal patterns. Later the ideas of BP were extended in different ways, being one particularly relevant for us the the application to the analysis of two dimensional images \cite{ribeiro2012complexity,zunino2016discriminating}.

For a time series $\mathcal{X}(t)$  the ordinal patterns procedure is based on the relative values of the neighbours belonging to the series, and in consequence takes into account the time structure or causality of the process that generated the sequence. To understand this idea, let us consider a real-valued discrete-time series $\mathcal{X}(t) = \{ x_t \in \mathbb{R} \}$, and the parameters $d \ge 2$ ( embedding dimension) and  $\tau \ge 1$ (the time delay). From these we construct a $d$-dimensional vector $\mathbf{Y}_t$:

\begin{equation}
\mathbf{Y}_{t} =  (x_{t-(d-1)\tau} , \dots , x_{t-\tau} , x_{t} )~ \qquad ~ {\mathrm {with}} ~\qquad ~ t \ge (d - 1)\tau  \ .
\end{equation}

The dynamical properties of the overall system is preserved by the vector $\mathbf{Y}_t$. We sorted in ascending order the components of the phase space trajectory $\mathbf{Y}_t$. Then, we can define a \textit{permutation vector}, $\mathbf{\Pi}_t$, with components given by the original position of the sorted values. Each one of these vectors represents a pattern (or motif) with $N_{\Pi}=d!$ possible patterns. For example let us consider the time series $\mathcal{X}(t)=\{0.42,~2.7,~4.2, ~0.35,~1.5 \}$ and take the parameters $d=3$ and $\tau=1$. We define the embedding vectors $\mathbf{Y}_t$ as $\mathbf{Y}_1=(0.42,~2.7,~4.2)$; $ \mathbf{Y}_2=(2.7,~4.2,~0.35)$; $\mathbf{Y}_3=(4.2,~0.35,~1.5)$, and the respective permutation vectors as $\mathbf{\Pi}_1=(0,~1,~2)$, $\mathbf{\Pi}_2=(1,~2,~0)$ and $\mathbf{\Pi}_3=(2,~0,~1)$.  Regarding the selection of the parameters, Bandt and Pompe \cite{Bandt2002} suggested working with $3 \leq d \leq 6$ and specifically consider an embedding delay $\tau = 1$. Nevertheless, other values of $\tau$ could provide additional information. It has been recently shown that this parameter is strongly related to the intrinsic time scales of the system under analysis \cite{zunino2010permutation, zunino2012distinguishing}. In addition, it is important  to take into account that for a reliable estimate of the ordinal probability distribution we need that, the length of the sequence $N_{\mathcal{X}}$ must to be greater than the number of possible ordinal patterns:  $N_{\Pi}<<N_{\mathcal{X}} $ \cite{Bandt2002}.

Ribeiro and Zunnino extended the permutation entropy framework to two-dimensional data \cite{ribeiro2012complexity,zunino2016discriminating}. We can consider a two-dimensional data array  $\{ x_u^v \}$ with $u=1,...,N_x$ and $v=1,...,N_y$ whose elements could be consider as pixels of an image. We define the embedding dimensions of the horizontal ($d_x$) and vertical  directions ($d_y$), and the corresponding $\tau_x$ and $\tau_y$. Using a similar scheme that in the time series case, we slice the data array in partitions of size $d_x \times d_y$ defined by:    

\begin{equation}
    \mathbf{Y}_i^j=\begin{bmatrix}
    x_i^j & x_{i}^{j+\tau_y} & \ldots  & x_{i}^{j+(d_y-1)\tau_y} \\ 
    x_{i+\tau_x}^j & x_{i+\tau_x}^{j+\tau_y} & \ldots  & x_{i + \tau_x}^{j+(d_y-1)\tau_y}\\ 
    \vdots & \vdots & \ddots & \\ 
    x_{i+(d_x-1)\tau_x}^j & x_{i+(d_x-1)\tau_x}^{j+\tau_y} & \ldots  & x_{i + (d_x-1)\tau_x}^{j+(d_y-1)\tau_y} 
    \end{bmatrix} 
\end{equation}

where $i = 1, \ldots , n_x$ and $j = 1, \ldots , n_y$ , with $n_x = N_x-(d_x-1)\tau_x$ and $n_y = N_y -(d_y-1)\tau_y$. 

To associate a permutation symbol with each two-dimensional partition we concatenate line by line the partition $y_i^j$ 
\begin{equation}
    \begin{split}
        \mathbf{Y}_i^j = & ( x_i^j, x_{i}^{j+\tau_y}, \ldots, x_{i}^{j+(d_y-1)\tau_y},\\
        & x_{i+\tau_x}^j, x_{i+\tau_x}^{j+\tau_y}, \ldots, x_{i + \tau_x}^{j+(d_y-1)\tau_y}\\
        & x_{i+(d_x-1)\tau_x}^j, x_{i+(d_x-1)\tau_x}^{j+\tau_y}, \ldots, x_{i + (d_x-1)\tau_x}^{j+(d_y-1)\tau_y}.)
    \end{split}
\end{equation}

Then we evaluate the permutation symbol associated with each data partition as in the one-dimensional case to define the symbolic array ${\mathbf{\Pi}_i^j }$ for $i=1,\ldots,n_x$ and $j=1,\ldots,n_y$ related to the data set. In this case the number of possible ordinal patterns are $N_{\Pi}=(d_xd_y)!$ \cite{ribeiro2012complexity}. For a better comprehension of the method we will present a simple example. Let us suppose we have the following  $3 \times 3$ array:
$$A=  \begin{bmatrix}
    2 & 3  & 7 \\ 
    4 & 5 & 6\\ 
    1 & 7  & 8 
    \end{bmatrix} $$

For this case we take $d_x=d_y=2$ and $\tau_x=\tau_y=1$ obtaining four partitions: $\mathbf{Y}_1=\begin{psmallmatrix} 2 & 3   \\  4 & 5   \end{psmallmatrix}=(2,3,4,5)$ with ordinal pattern $\mathbf{\Pi}_1=(0,1,2,3)$; $\mathbf{Y}_2=\begin{psmallmatrix} 3 & 7   \\  5 & 6   \end{psmallmatrix}=(3,7,5,6)$ with ordinal pattern $\mathbf{\Pi}_2=(0,3,1,2)$;
$\mathbf{Y}_3=\begin{psmallmatrix} 4 & 5 \\  1 & 7   \end{psmallmatrix}=(4,5,1,7)$ with ordinal pattern $\mathbf{\Pi}_3=(1,2,0,3)$ and
$\mathbf{Y}_4=\begin{psmallmatrix} 5 & 6   \\  7 & 8   \end{psmallmatrix}=(5,6,7,8)$ with ordinal pattern $\mathbf{\Pi}_4=(0,1,2,3)$.

Once we have all the ordinal patters $\mathbf{\Pi}$ corresponding to a signal or a image we compute the probability distribution  $P_{op}=P(\Pi_i)$ with $i=1,\ldots,d!$ for one-dimensional sequence or $i=1,\dots,(d_xd_y)!$ for two-dimensional arrays. 

We will compare two signals or images by evaluating a divergence between the corresponding probability distributions for the associated ordinal patterns, $P_{op}$ and $Q_{op}$. We used the BRC, that we name the $\gamma$-divergence, explicitly given by:

\begin{equation}\label{Definicion formal}
\mathcal{D}_{\gamma}(P_{op}||Q_{op})= 2\mathop{\sum}_{i=1}^{N_{\Pi}}\gamma_{g}(p(\Pi_i),q(\Pi_i))
\end{equation}
with
\begin{equation}
\gamma_{g}(p(\Pi_i),q(\Pi_i))= \frac{1}{2}p(\Pi_i) ~ g(p(\Pi_i)) + \frac{1}{2} q(\Pi_i) ~ g(q(\Pi_i))- \frac{1}{2} (p(\Pi_i)+q(\Pi_i)) ~ g\left(\frac{p(\Pi_i)+q(\Pi_i)}{2}\right)
\end{equation}
The only requirement on the function $g(x)$ is that the sum $\sum_i x_i g(x_i)$ results concave.

\section{Applications}
\label{sec:application}
To investigate the behaviour of the $\gamma$-family we use different $\gamma$-functions, $xg(x)$, by changing the $g$ function. We apply them to four cases. In the first example we analyse the coupling of the Henon-Henon system; in the second one we detect the change in the dynamics between chaotic and stochastic signals; in the third example we apply our scheme to detect the changes in the stages of an EEG signal and finally we evaluate the distances between textured 2D images.  In each one of this examples, we work with four $g(x)$ function; i) $g(x)=e^x$; $g(x)= ~log(x)$; $g(x)=~\sqrt[]{x}$ and $g(x)=~ sinh(x)$. 

\subsection{Henon-Henon coupled system }

The coupled Henon-Henon system is described by the following set of equations \cite{paluvs2001synchronization,krakovska2015causality}:

\begin{align*} 
y_1(n+1) &=  1.4-y_1^2(n)+b~y_2(n) \\ 
y_2(n+1) &=  y_1(n)\\
x_1(n+1) &= 1.4 - (\epsilon~ y_1(n) + (1-\epsilon)~x_1(n)+b~x_2(n))\\
x_2(n+1) &= x_1(n)
\end{align*}
where, in our choice $b = 0.3$. For the simulation the coupling parameter $\epsilon$ was varied from zero to $1$ in steps of $0.1$. For each $\epsilon$, $200$ realisations were computed using random initial conditions. The signals length were $N=100000$ and the ordinal patterns were evaluated with the parameters $d=3,4,5$ and $\tau=1$. 

\begin{figure}[htbp]
\centerline{\includegraphics[width=\textwidth]{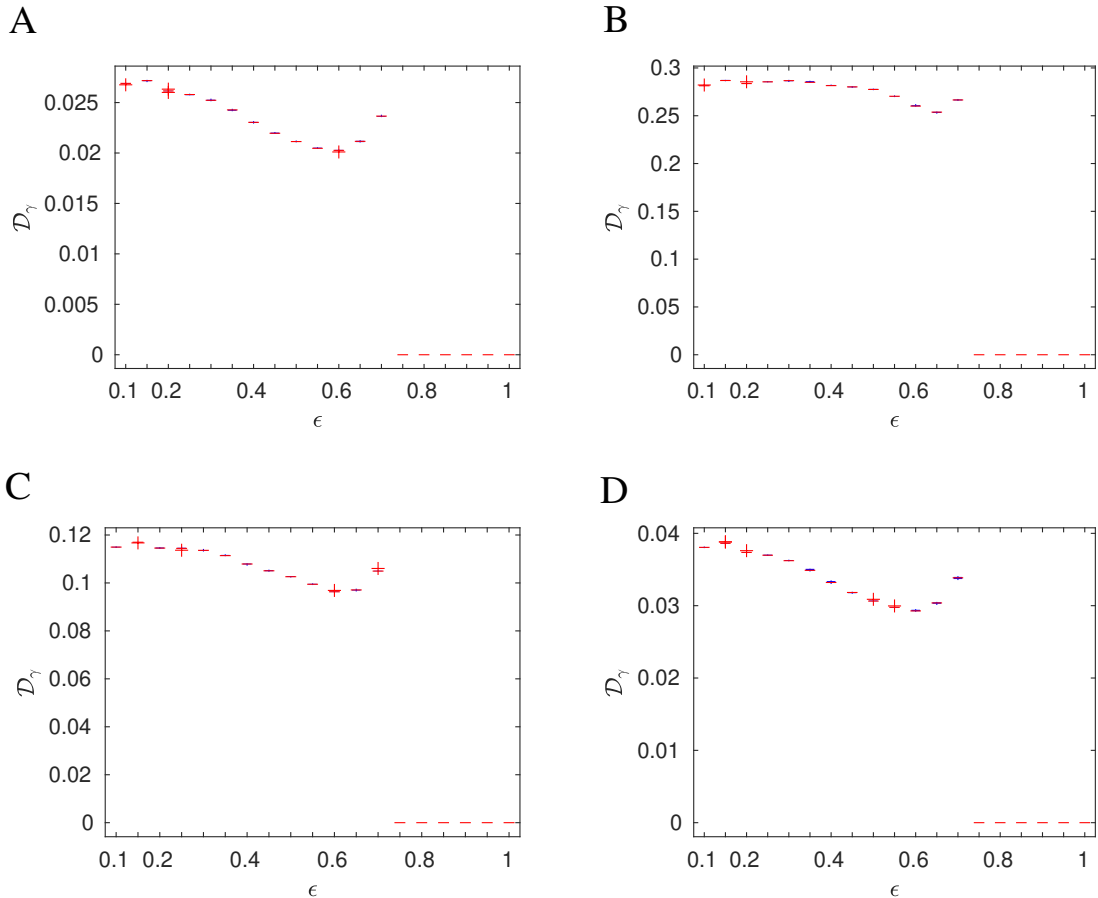}}
\caption{ Henon-Henon coupled system. Boxplot  $\gamma$-divergence  as a function of the coupling parameter $\epsilon$. $\Dg$ was calculated  with function $g(x)$ A) $e^x$, B) $log(x)$, C) $\sqrt[]{x}$, D) $sinh(x)$. The ordinal patterns parameter used were $d = 4$ and $\tau = 1$.}
\label{Fig:Hennon-Hennon}
\end{figure}

The Figure \ref{Fig:Hennon-Hennon} shows the results for the four divergences as a function of the coupling parameter $\epsilon$. The median value of estimator $\Dg$ is maximum for $\epsilon = 0$ in all the cases. This is expected because there is no information sharing between the two systems. The $\Dg$ values decrease with the coupling parameter up to the value $\epsilon = 0.6$. In contrast, for $\epsilon \geq 0.7$ the  $\Dg$ values are zero. For these values of the coupling parameter, the Henon-Henon system is synchronised in such a way that both systems are statistically indistinguishable. Analogous results were obtained using the parameter $d=3,5$.

Similar behaviour has been already observed on other information quantifiers such as the transfer entropy \cite{paluvs2001synchronization,krakovska2015causality,restrepo2020transfer}.
It is important to note that for the function $g(x)=log(x)$ (that is, the JSD) the values between the maximum and zero is $ \sim 0.3$ almost an order of magnitude higher than for the other functions $g(x)$.
 
\subsection{Dynamical changes detection }

\begin{figure}[htbp]
\centerline{\includegraphics[width=\textwidth]{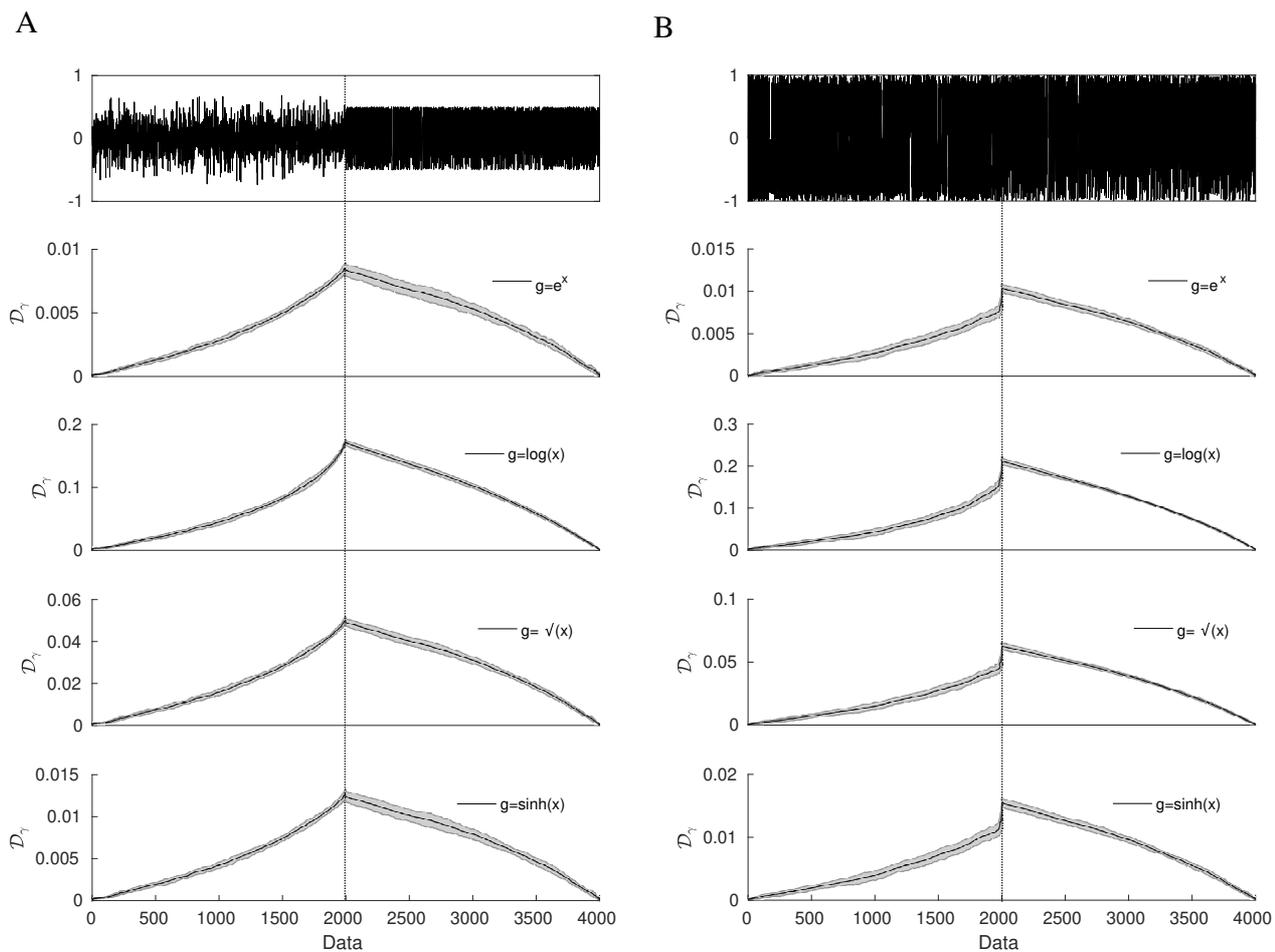}}
\caption{Dynamical changes detection in mixed signal using $\gamma$-divergence for different function $g(x)$. A) Mixed signal white noise - logistic map, B) Mixed signal cubic map - logistic map. 
The ordinal patterns parameter used were $d = 4$ and $\tau = 1$.}
\label{Fig:Chaos-Niose}
\end{figure}

In this example we use the $\gamma$-divergence to detect changes in the dynamics of a signal. For this, we generated a group of mixed signals: stochastic-chaotic and chaotic-chaotic. For the chaotic signal we use the  \textit{logistic map} \cite{may2004simple} given by the equation $x_{n+1}=r~x_{n}\left(1-x_{n}\right)$ for $r=4$ and the \textit{cubic map} \cite{wan1985symbolic} $x_{n+1}=A~x_{n}\left(1-x_{n}^2\right) $ for $A=3$. For stochastic signal we use a \textit{white noise}.   
We generate  $N=100$ mixed signals $S_{s-c}^i=S_s+S_c=2000+2000=4000$ and $S_{c_1-c_2}^i=S_{c_1}+S_{c_2}=2000+2000=4000$ with $i=1,...,100$ (Figure \ref{Fig:Chaos-Niose} top). For each signal we used a pointer method to study the divergence. This method consists of a sliding pointer $p$ which move point to point over the signal in the range $2d! \leq i \leq L-2d! $ splitting the signals in two subsequences $S_l[x(1),\ldots,x(p)]$ and $S_r[x(p+1),\ldots,x(L)]$. In this case we apply the weighted divergence define in eq.\ref{weighted}  with weight $w_l=L_{S_l}/L$ and $w_r=L_{S_r}/L$. We compute the $\Dg^{w_l,w_r}(S^l||S^r)$ for all $i$-signal and finally measure the mean value $\mu=<\Dg>_i$ and the standard deviation $\sigma$. 
In figure \ref{Fig:Chaos-Niose} are shown the results for the two  studied cases: white noise-logistic map signal (A) and cubic-logistic maps signal (B). The figure shows the $\mu$ (straight line) and $\sigma$ (shadow bar) for the four above indicated $g(x)$ functions. I both cases the maximum value ($\Dg^{max}$) of the divergence is reached in the transition between the two signals. Is important to remark that for every function $g(x)$ the method detects the change in the dynamics but the highest values of $\Dg^{max}$ corresponds to $g(x)=log(x)$. 

\subsection{Detection of different states of sleep}

In this example we use the proposed scheme to identify the transition between sleep states from an electroencephalogram (EEG) record. 

The problems associated with sleep have serious repercussions on people's health. That is why their study is of great clinical relevance. \cite{benca1997sleep,stickgold2005sleep,pilcher1997sleep}. 

Usually two primary sleep stages are identified: \textit{rapid eye movement} (REM) sleep and \textit{non-REM} sleep (NREM). REM is defined as an active sleeping period showing an intense brain activity. Brain waves remain fast and desynchronised similar to those in the waking state. In NREM sleep stage the physiological activity decrease, the brain waves get slower and have greater amplitude, breathing, and heart rate slows down and blood pressure drops. The NREM phase is composed by three stages: N1, N2, and N3. The N1 stage is characterised by perceived drowsiness or transition from being awake to falling asleep observed by slowing down the brain waves and muscle activity. Stage N2 is a period of light sleep during which eye movement stops. Brain waves become slower (Theta waves  (4-7 Hz)) with occasional bursts of rapid waves (12-14 Hz) called \textit{sleep spindles}, coupled with spontaneous periods of muscle tone mixed. Lastly, stage N3 is characterised by the presence of Delta (0.5-4 Hz) slow waves, interspersed with smaller, faster waves \cite{boostani2017comparative}. The N3 stage is a deep sleep stage, without eye movement, and a decrease of muscle activity, resembling a coma state. Usually, sleepers pass through these four stages (REM, N1, N2, and N3) cyclically. A complete sleep cycle takes an average of 90 to 110 minutes, with each one lasting between 5 to 15 minutes. These cycles must be maintained for healthy body function in awake state \cite{Ogilvie2001}.
Developing tools that can detect the changes in dynamic sleep stages over EEG signal are highly essential for studying  patients with sleep disorders \cite{tononi2006sleep}. Here we use our  $\gamma$-divergence to detect changes in the dynamics of the EEG to  detect the transition from  one state to another.

\begin{figure}[htbp]
\centerline{\includegraphics[width=\textwidth]{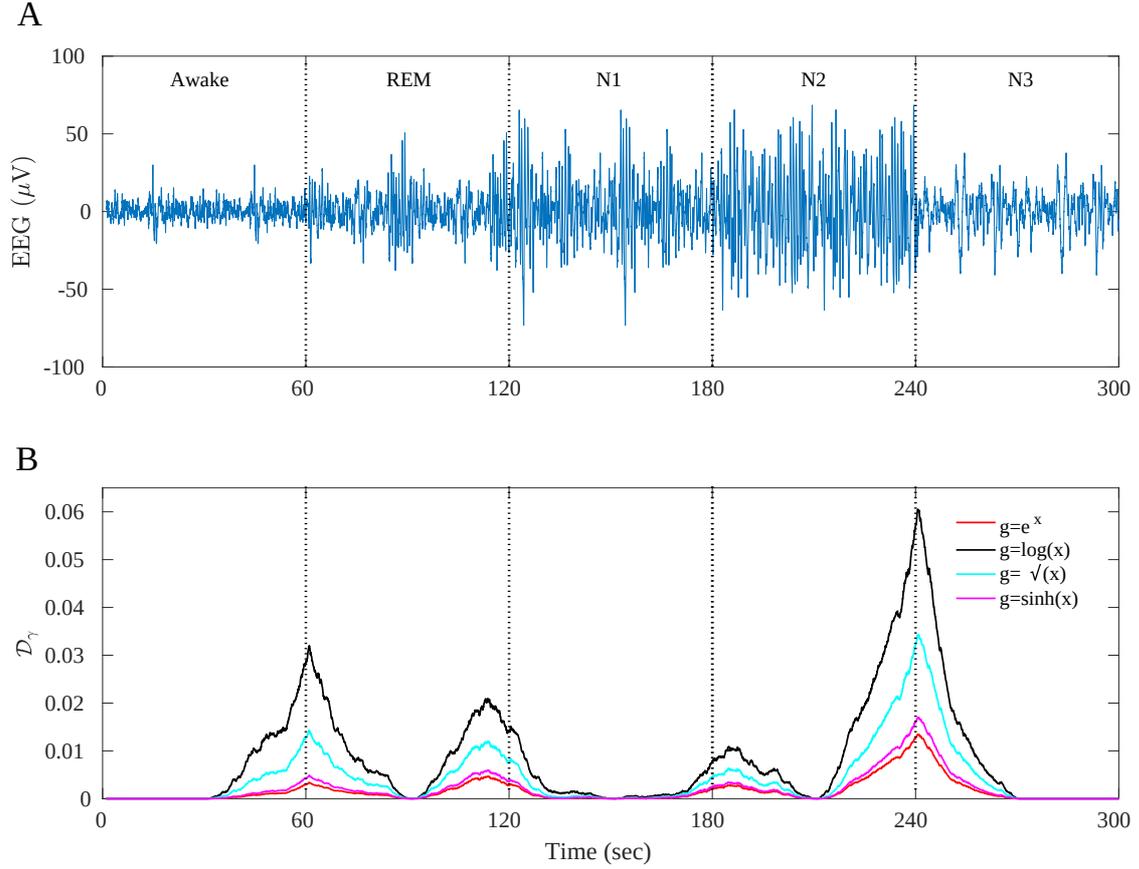}}
\caption{ Application of the $\gamma$-divergence over a sleep EEG signal using a running windows. A) The EEG signal is composed by $5$ sub-signal belonging to different sleep stages (Awake, REM, N1, N2, N3). Each state has $6000$ point corresponding to $60$ sec recording  (dashes horizontal lines). B) Results corresponding to the running windows method (see text for details). The signal was quantified with the permutation vectors with parameter $d=4$ and $\tau=1$. For all functions, the maximum values $\D_{\gamma_{max}}$  were reached in the exact point where a transition between sleep states exists.}
\label{Fig3}
\end{figure}

The data were taken from the \textit{Physionet database:  The Sleep-EDF Database [Expanded]} \cite{goldberger1997physiobank,kemp2000analysis}, and are freely available at \cite{Physionet}. The EEG were recorded (Fpz-Cz) thorough  bipolar channels and the sampling frequency was $100$ Hz. Initially, we extracted segments from the original signal belonging to the five  different sleep states: Awake, REM, N1, N2, N3 \footnote{We used the notes provided by the database}. Each segments had $6000$ points (corresponding to $60$ sec of recording) and were joined into a single signal \footnote{This was done for a better visualisation of the results, because the time of each sleep state can vary from seconds to several minutes.}  (Figure \ref{Fig3}A.) The signal was preprocessed with a band-pass filter between $0.5-60$ Hz. The ordinal pattern were computed using the parameter $d=4$ and $\tau=1$. 

Figure \ref{Fig3}B shows that our method could detect the transition between different sleep states  for all the used functions $g(x)$. The $\log(x)$  function shows the highest values and the best differentiation between stages. The transition between Awake-REM and N2-N3 are more remarkable than in REM-N1 and N1-N2. There is a significant differentiation between N2-N3 because of N3 is the deepest sleep stage. In this stage the body becomes more insensitive to outside stimuli. In this state, the EEG signal is mostly composed by slow waves (Delta and Theta) causing  the brain dynamics to be sharply different from the other states. Lower $\gamma$-divergences values  were found between N1-N2 states showing that both states share similar characteristics in their dynamics. Particularly, N1 is characterised by drowsiness slowing down the brain waves and muscle activity. N1-N2) are very similar stages being the main difference that in N1 occasional bursts of rapid waves (12-14 Hz), called \textit{sleep spindles} appear.

Similar result can be found between REM and N1. REM and N1 also present lower values of the gamma divergence. This is something to be expected considering that the EEG during the REM stage contains frequencies present in the ``awake'' state and in the lighter stages of sleep N1 \cite {nicolaou2011use}. Despite the similarities of the REM and N1, there are still enough differences between them such that statistically different values are obtained. The presence of $11-16$ Hz activity (sleep spindles) in N1, and more abundant alpha activity ($8-13$Hz) in REM sleep means that these two stages present activity at an overlapping frequency range, which explains the proximity of the divergence values obtained. Difficulty in detecting N1 and REM sleep has also been found using other measures \cite {noirhomme2009bispectral, mateos2021using}.


\subsection{Distance between textured images}

As a last application we use the $\gamma$-divergence to measure the distance between 2D textured images.  
The images were taken from the Normalized Brodatz Texture (NBT) album avalible in
\href{https://multibandtexture.recherche.usherbrooke.ca}{https://multibandtexture.recherche.usherbrooke.ca}. 
This normalised database has 112 texture images an improvement regarding the original Brodatz texture database since that grayscale background effects have been removed \cite{abdelmounaime2013new}. 

\begin{figure}[htbp]
\centerline{\includegraphics[width=\textwidth]{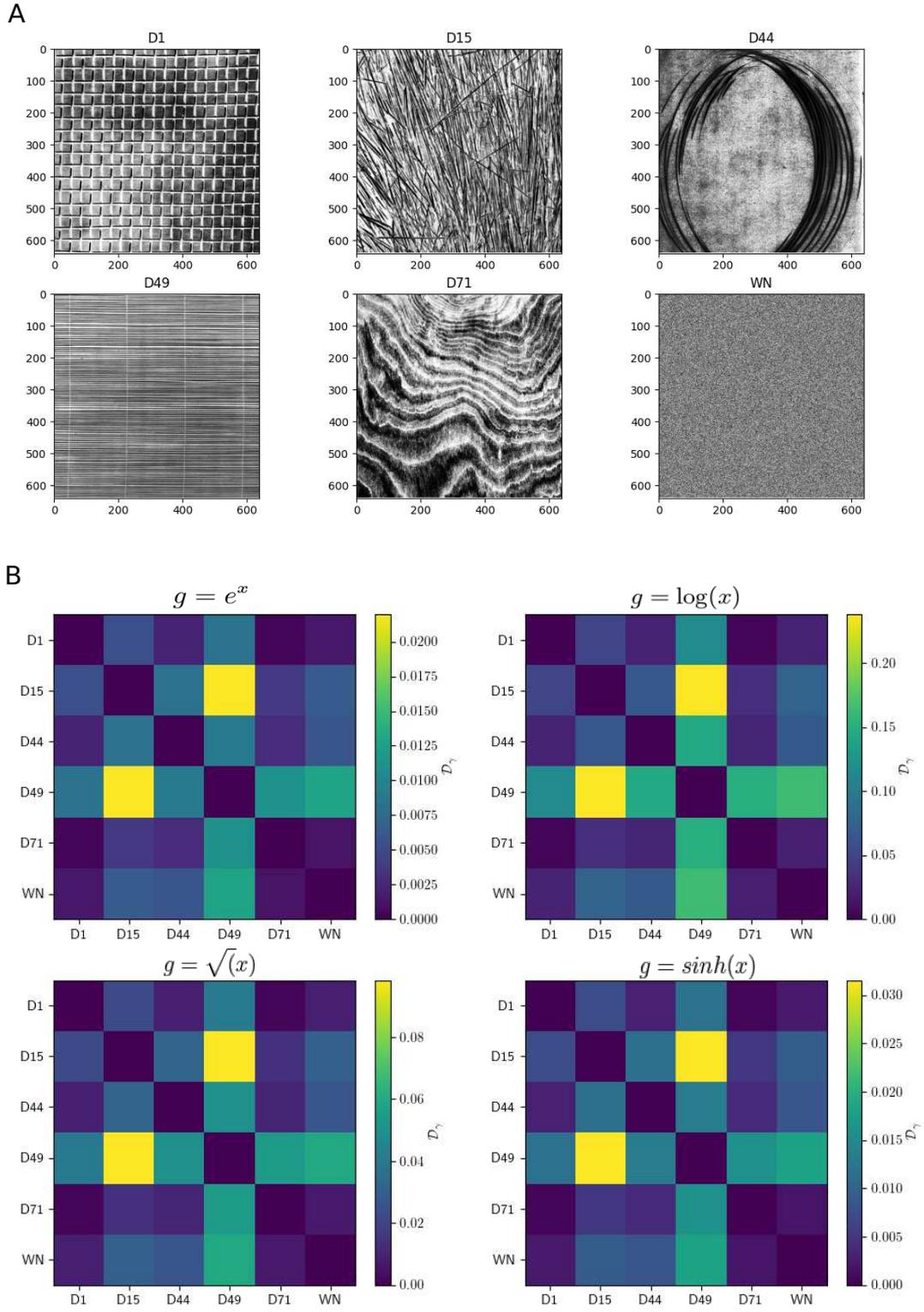}}
\caption{ A) Six samples of the NBT album with their corresponding labels are illustrated. B) Location of the 112 texture images from the NBT album in the CECP. $\gamma$-divergence matrix between textured images showing in A 
using 2D ordinal patterns with $d_x = d_y  = 2$ and $\tau_x = \tau_y = 1$.  }
\label{Fig:Texture_images}
\end{figure}

Because of this, it is impossible to discriminate between textures from this normalised database using only ﬁrst order statistics. Six different samples of the NBT database are illustrated in Figure \ref{Fig:Texture_images}A. The images of the NBT album have dimensions of $640 \times 640$ pixels and 8 bits/pixel, which provides 256 grayscale levels. 

For the 2D quantization, the ordinal parameters used were $d_x = d_y  = 2$ and $\tau_x = \tau_y = 1$. Figure \ref{Fig:Texture_images}B shows the divergence matrices calculated for the different functions $g(x)$. We can see all the divergence values allow us to distinguish different textures from each other. The divergence is higher when the texture patterns are more different, for example between D15 and D47, and lower when the patterns have a similar geometry for example between D1 and D71. The four divergences show similar relative values between the images however, the values for $g(x)=log(x)$ have an order of magnitude higher than the other functions.

\section{Discussion}
\label{sec:Discussion}
We showed some particular aspects of the JSD when thought of as a BRC. In particular that is the only BRC that is a Csiszar divergence besides being one of the few that has a metric character. Then we defined the weighted BRC, which can be explored in the context of Bayesian inference. We restrict ourselves to the application of these weighted version to the analysis of the dynamical behaviour of time series.  Afterwards, we showed that the ordinal patterns method is a natural way to apply divergence in signal and image analysis. The ordinal patterns allow a natural quantification for the use of one special kind of BRC that we called $\gamma$-divergence.  

We applied the $\gamma$-divergence to four examples, three for signals analysis and one for image analysis. Among the signal analyses, we studied the coupling between two signals generated by the Hennon-Hennon system. We observed that divergence can quantify the coupling of the system depending on the coupling parameter. In the second example, we use divergence to detect changes in the dynamics of combined signals. We noted that divergence can accurately detect the transition point between two chaotic-stochastic and chaotic-chaotic systems. Later, we analysed EEG signal from sleep patient. We could detect the points where the signal changes its dynamics because of the change in the state of sleep. Showing it can be an alternative tool for detecting different sleep states. Finally, we used this family of divergences to study distances between textured images. We could  realised divergence values were smaller for similar textures and larger for different textures. 

One interesting result is for all the examples studied, the largest divergence values occur for the generating function $g(x)=log(x)$ which is precisely the Jensen-Shannon divergence.



It is also important to mention that when using divergences, we had consider the significance of the value obtained. This significance is what allows us to say if the distance found is really a real and not just a measure of the statistical fluctuations. In this work we used the standard deviation  calculated over $N$ realizations as a measure of significance. However, many times we do not have more than one realization being this method useless.  Therefore, a  theoretical study of the field is necessary. Some studies on this topic have already been addressed by Grosse et al. for the Jensen-Shannon divergence \cite{grosse2002analysis}. Similar studies should be carried out for this family of  $\gamma$-divergence. However, this exceeds this work but will be a addressed in the near future.

\section{Aknowledgements} We are grateful to Secretaria de Ciencia y Tecnica de la Universidad Nacional de C\'ordoba, Argentina, for financial assistance. This work was partially supported by Grant PIP 519/2019 from CONICET, Argentina and  by the Grant PICT 2019-01750 from Angencia Nacional de Promoci\'on de la Investigaci\'on, el Desarrollo Tecnol\'ogico y la Inovaci\'on, Argentina.


\bibliography{Gamma_bibliografia.bib}

\begin{thebibliography}{10}

\bibitem{fisher1925theory}
R.~A. Fisher.
\newblock Theory of statistical estimation.
\newblock In {\em Mathematical Proceedings of the Cambridge Philosophical
  Society}, volume~22, pages 700--725. Cambridge University Press, 1925.

\bibitem{cover2012elements}
T.~M. Cover and J.~A. Thomas.
\newblock {\em Elements of information theory}.
\newblock John Wiley \& Sons, 2012.

\bibitem{parry2000probability}
M.~Parry and E.~Fischbach.
\newblock Probability distribution of distance in a uniform ellipsoid: Theory
  and applications to physics.
\newblock {\em Journal of Mathematical Physics}, 41(4):2417--2433, 2000.

\bibitem{ondimu2008effect}
S.~N. Ondimu and H.~Murase.
\newblock Effect of probability-distance based markovian texture extraction on
  discrimination in biological imaging.
\newblock {\em Computers and Electronics in Agriculture}, 63(1):2--12, 2008.

\bibitem{kostin2010probability}
A.~Kostin.
\newblock Probability distribution of distance between pairs of nearest
  stations in wireless network.
\newblock {\em Electronics Letters}, 46(18):1299--1300, 2010.

\bibitem{majtey2005jensen}
A.~P. Majtey, P.~W. Lamberti, and D.~P. Prato.
\newblock Jensen-shannon divergence as a measure of distinguishability between
  mixed quantum states.
\newblock {\em Physical Review A}, 72(5):052310, 2005.

\bibitem{crooks}
G.~Crooks and D.~Sivak.
\newblock Measures of trajectory ensemble disparity in nonequilibrium
  statistical dynamics.
\newblock {\em Journal of Statistical Mechanics: Theory of Experiments},
  P06003, 2011.

\bibitem{osan2018monoparametric}
T.~M Os{\'a}n, D.~G. Bussandri, and P.~W. Lamberti.
\newblock Monoparametric family of metrics derived from classical
  jensen--shannon divergence.
\newblock {\em Physica A: Statistical Mechanics and its Applications},
  495:336--344, 2018.

\bibitem{majtey2004monoparametric}
A.~P. Majtey, P.~W. Lamberti, and A.~Plastino.
\newblock A monoparametric family of metrics for statistical mechanics.
\newblock {\em Physica A: Statistical Mechanics and its Applications},
  344(3-4):547--553, 2004.

\bibitem{csiszar1974information}
I.~Csisz{\'a}r.
\newblock Information measures: A critical survey.
\newblock In {\em Transactions of the Seventh Prague Conference on Information
  Theory, Statistical Decision Functions, Random Processes}, pages 73--86,
  1974.

\bibitem{burbea1982entropy}
J.~Burbea and C.~R. Rao.
\newblock Entropy differential metric, distance and divergence measures in
  probability spaces: A unified approach.
\newblock {\em Journal of Multivariate Analysis}, 12(4):575--596, 1982.

\bibitem{ali1966general}
S.~M. Ali and S.~D. Silvey.
\newblock A general class of coefficients of divergence of one distribution
  from another.
\newblock {\em Journal of the Royal Statistical Society: Series B
  (Methodological)}, 28(1):131--142, 1966.

\bibitem{Bandt2002}
C.~Bandt and B.~Pompe.
\newblock {Permutation entropy: a natural complexity measure for time series.}
\newblock {\em Physical Review Letters}, 88(17):174102, 2002.

\bibitem{zanin2021ordinal}
Massimiliano Zanin and Felipe Olivares.
\newblock Ordinal patterns-based methodologies for distinguishing chaos from
  noise in discrete time series.
\newblock {\em Communications Physics}, 4(1):1--14, 2021.

\bibitem{ribeiro2012complexity}
H.~V. Ribeiro, L.~Zunino, E.~K. Lenzi, P.~A. Santoro, and R.~S. Mendes.
\newblock Complexity-entropy causality plane as a complexity measure for
  two-dimensional patterns.
\newblock 2012.

\bibitem{zunino2016discriminating}
L.~Zunino and H.~V. Ribeiro.
\newblock Discriminating image textures with the multiscale two-dimensional
  complexity-entropy causality plane.
\newblock {\em Chaos, Solitons \& Fractals}, 91:679--688, 2016.

\bibitem{zunino2010permutation}
L.~Zunino, M.~C. Soriano, I~Fischer, O.~A. Rosso, and C.~R. Mirasso.
\newblock {Permutation-information-theory approach to unveil delay dynamics
  from time-series analysis}.
\newblock {\em Physical Review E}, 82(4):46212, 2010.

\bibitem{zunino2012distinguishing}
L.~Zunino, M.~C. Soriano, and O.~A. Rosso.
\newblock {Distinguishing chaotic and stochastic dynamics from time series by
  using a multiscale symbolic approach}.
\newblock {\em Physical Review E}, 86(4):46210, 2012.

\bibitem{paluvs2001synchronization}
V.~Palu{\v{s}}, M.and~Kom{\'a}rek, Z.~Hrn{\v{c}}{\'\i}{\v{r}}, and
  K.~{\v{S}}t{\v{e}}rbov{\'a}.
\newblock Synchronization as adjustment of information rates: Detection from
  bivariate time series.
\newblock {\em Physical Review E}, 63(4):046211, 2001.

\bibitem{krakovska2015causality}
A.~Krakovsk{\'a}, J.~Jakub{\'\i}k, H.~Bud{\'a}{\v{c}}ov{\'a}, and
  M.~Holecyov{\'a}.
\newblock Causality studied in reconstructed state space. examples of
  uni-directionally connected chaotic systems.
\newblock {\em arXiv preprint arXiv:1511.00505}, 2015.

\bibitem{restrepo2020transfer}
J.~F. Restrepo, D.~M. Mateos, and G.~Schlotthauer.
\newblock Transfer entropy rate through lempel-ziv complexity.
\newblock {\em Physical Review E}, 101(5):052117, 2020.

\bibitem{may2004simple}
R.~M. May.
\newblock Simple mathematical models with very complicated dynamics.
\newblock In {\em The Theory of Chaotic Attractors}, pages 85--93. Springer,
  2004.

\bibitem{wan1985symbolic}
Z.~Wan-zhen, D.~Ming-zhou, and L.~Jia-nan.
\newblock Symbolic description of periodic windows in the antisymmetric cubic
  map.
\newblock {\em Chinese Physics Letters}, 2(7):293, 1985.

\bibitem{benca1997sleep}
R.~M. Benca, M.~Okawa, M.~Uchiyama, S.~Ozaki, T.~Nakajima, K.~Shibui, and W.~H.
  Obermeyer.
\newblock Sleep and mood disorders.
\newblock {\em Sleep medicine reviews}, 1(1):45--56, 1997.

\bibitem{stickgold2005sleep}
R.~Stickgold.
\newblock Sleep-dependent memory consolidation.
\newblock {\em Nature}, 437(7063):1272--1278, 2005.

\bibitem{pilcher1997sleep}
J.~J. Pilcher and A.~S. Walters.
\newblock How sleep deprivation affects psychological variables related to
  college students' cognitive performance.
\newblock {\em Journal of American College Health}, 46(3):121--126, 1997.

\bibitem{boostani2017comparative}
R.~Boostani, F.~Karimzadeh, and M.~Nami.
\newblock {A comparative review on sleep stage classification methods in
  patients and healthy individuals}.
\newblock {\em Computer methods and programs in biomedicine}, 140:77--91, 2017.

\bibitem{Ogilvie2001}
R.~D. Ogilvie.
\newblock {The process of falling asleep}.
\newblock {\em Sleep Medicine Reviews}, 5(3):247--270, 2001.

\bibitem{tononi2006sleep}
G.~Tononi and C.~Cirelli.
\newblock Sleep function and synaptic homeostasis.
\newblock {\em Sleep medicine reviews}, 10(1):49--62, 2006.

\bibitem{goldberger1997physiobank}
A.~L. Goldberger, L.~Amaral, L.~Glass, J.~M. Hausdorff, P.~C. Ivanov, R.~G.
  Mark, J.~E. Mietus, G.~B. Moody, C.~K. Peng, and H.~E. Stanley.
\newblock Physiobank, physiotoolkit, and physionet: Circulation.
\newblock {\em Discovery}, 101(23):1, 1997.

\bibitem{kemp2000analysis}
B.~Kemp, A.~H. Zwinderman, B.~Tuk, H.~A.~C. Kamphuisen, and J.~J.~L. Oberye.
\newblock Analysis of a sleep-dependent neuronal feedback loop: the slow-wave
  microcontinuity of the eeg.
\newblock {\em IEEE Transactions on Biomedical Engineering}, 47(9):1185--1194,
  2000.

\bibitem{Physionet}
Physionet databanck.
\newblock {Sleep-EDF Database Expanded}.
\newblock
  $\backslash$url{\{}https://physionet.org/content/sleep-edfx/1.0.0/{\}}.

\bibitem{nicolaou2011use}
N.~Nicolaou and J.~Georgiou.
\newblock {The use of permutation entropy to characterize sleep
  electroencephalograms}.
\newblock {\em Clinical EEG and Neuroscience}, 42(1):24--28, 2011.

\bibitem{noirhomme2009bispectral}
Q.~Noirhomme, M.~Boly, V.~Bonhomme, P.~Boveroux, C.~Phillips, P.~Peigneux, and
  Others.
\newblock Bispectral index correlate with regional cerebral blood flow during
  sleep.
\newblock {\em Archives Italiennes de Biologie}, 147(1/2):51--57, 2009.

\bibitem{abdelmounaime2013new}
S.~Abdelmounaime and H.~Dong-Chen.
\newblock New brodatz-based image databases for grayscale color and multiband
  texture analysis.
\newblock {\em International Scholarly Research Notices}, 2013, 2013.

\bibitem{grosse2002analysis}
I.~Grosse, P.~Bernaola-Galv{\'a}n, P.~Carpena, R.~Rom{\'a}n-Rold{\'a}n,
  J.~Oliver, and H.E. Stanley.
\newblock Analysis of symbolic sequences using the jensen-shannon divergence.
\newblock {\em Physical Review E}, 65(4):041905, 2002.

\end{thebibliography}
\bibliographystyle{unsrt}

\end{document}